\documentclass[floatfix,prb,twocolumn,amsmath,amssymb]{revtex4}

\usepackage{color}

\newcommand{\fref}[1]{{figure~\ref{#1}}}

\newcommand{\sref}[1]{{section~\ref{#1}}} 

\usepackage{setspace}
\usepackage{subfigure}
\usepackage{graphicx, epsfig}
\usepackage{dcolumn}
\def\be{\begin{equation}}
\def\ee{\end{equation}}
\def\bea{\begin{eqnarray}}
\def\eea{\end{eqnarray}}
\begin{document}

\title{Ground-state properties of the attractive one-dimensional Bose-Hubbard model}

\author{Norman Oelkers}
		\email{oelkers@maths.uq.edu.au}
\author{Jon Links}
		\email{jrl@maths.uq.edu.au}
\affiliation{Centre for Mathematical Physics, The University of Queensland, Brisbane 4072, Australia}

\pacs{XXX}

\date{\today}
\begin{abstract}
We study the ground state of the attractive one-dimensional Bose-Hubbard model, 
and in particular the nature of the crossover between the weak interaction and strong interaction regimes
for finite system sizes. Indicator properties like the gap between the ground and first excited energy levels, 
and the incremental ground-state wavefunction overlaps are used to locate different regimes.
Using mean-field theory we predict that there are two distinct crossovers 
connected to spontaneous symmetry breaking of the ground state. 
The first crossover arises in an analysis valid for large $L$ with
finite $N$, where $L$ is the number of lattice sites and $N$ is the total particle number. 
An alternative  approach valid for large $N$ with finite $L$ yields a second crossover. 
For small system sizes we numerically investigate the model
and observe that there are signatures of both crossovers. 
We compare with exact results from Bethe ansatz methods in several limiting cases to explore
the validity for these numerical and mean-field schemes.
The results indicate that for finite attractive systems there are 
generically three ground-state phases of the model.  
\end{abstract}
\maketitle
\section{Introduction}
\label{secI}
\begin {figure}[t]
\begin{minipage}{\linewidth}
\begin{center}
\includegraphics[angle=0,width=0.75\linewidth]{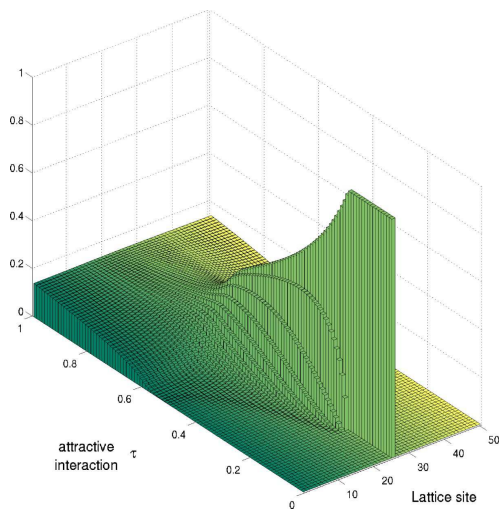}
\end{center}
\end{minipage}
\caption{
Generic ground-state behaviour for attractive bosons: momentum distribution of
trapped bosons, within a mean-field approach for $L=50$.
See also \fref{fig:MFT-real-dist} for the real space ``density`` and section \ref{sec:Generic-MFT} for technical details.
In the weak interaction region (right front in these figures: $\tau\to0$) the system is in an ideal 
BEC state, all bosons are condensed
into the lowest momentum and the semi-classical density is flat. For strongly interacting bosons 
(left rear in these figures: $\tau\to1$)
the momentum distribution is flat.
while
the translational symmetry in real space is broken, all bosons are on the same side.
In between there is a rich cross over regime which we study in this paper.
}
\label{fig:MFT-mom-dist}
\end{figure}

\begin {figure}[t!]
\begin{minipage}{\linewidth}
\begin{center}
\includegraphics[angle=0,width=0.72\linewidth]{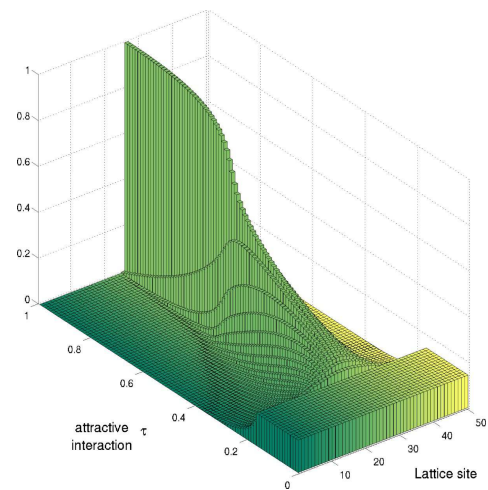}
\end{center}
\end{minipage}
	\caption{
	Real space density for attractive bosons in a trap with $L=50$ sites in a semi-classical theory.
	This picture is the corresponding density to \fref{fig:MFT-mom-dist},
	see  caption for different physical regimes.
	Technical details text are in section \ref{sec:Generic-MFT}.
	}
\label{fig:MFT-real-dist}
\end{figure}

%  \subsubsection{General area of interest}
Systems of attractive bosons are one of the most intriguing current topics in physics.
For instance they 
might lead the way for fabricating mesoscopic Schr\"odinger cat states \cite{Leggett, Zoller},
and in the experimental context, they have been used to produce the {\it Bose\-nova} phenomena \cite{bosenova}.   
The substantial amount of research undertaken recently~\cite{  
Hulet,
Salomon,
attractive-through-Feshbach,
Kanamoto,
attractive-BH-numerics,
Buonsante,
Pan,
attractive-integrable-LL-model} poses new questions surrounding systems of attractive bosons.
%  \subsubsection{Experiments}
An almost ideal realisation of a lattice Bose gas - 
the Bose-Hubbard model - has been found
in bosons trapped inside optical lattices~\cite{BH-optical-lattice-realisations}. 
The use of techniques like Fesh\-bach resonances allows tuning of the scattering length, i.e. changing the interaction strengh,
even crossing from repulsive to attractive~\cite{attractive-through-Feshbach,Salomon}.
The theoretical boson model predicts a dramatic change in the ground state
of a large but finite system
when the attractive interaction
strength is varied from weak to strong attractive, see 
\fref{fig:MFT-mom-dist} and \fref{fig:MFT-real-dist}
for visualisation.
Currently technical difficulties make experiments on attractive system
considerably harder compared to repulsive system~\cite{Hulet,Salomon}.
Once handling and stability of attractive bosons in optical lattices allows experiments at controlled and varying interaction strengths
this general transitional feature could be experimental verifiable.
Potential candidates for measurements might include correlation functions, momentum distribution after release from the trap
and the low-lying energy spectrum.
% \subsubsection{Previous theoretical work}
Historically, the theoretical study of attractive bosonic systems has received little attention due to difficulties~\cite{Mattis-red-book}
like non-saturation or high site occupancy. 
A number of numerical and approximative studies for a variety of attractive bosonic 
systems~\cite{Kanamoto,attractive-BH-numerics,Buonsante,Pan} have found 
a transitional regime between the strong and the weak interacting regions.
This crossover can be seen in properties like
the energy spectrum~\cite{Kanamoto}, correlation functions~\cite{Buonsante} or entanglement~\cite{Pan}. 
All these properties have the common feature  
that the crossover becomes sharper and more pronounced for larger system sizes,
a region where numerical and approximative
techniques enter a region of uncertainty.
A transition is also seen in studies of mean-field techniques of the non-linear Schr\"odinger equation
in the context of the Bogolyobov approximation or in solitonic solutions of the Gross-Pitaevskii equation~\cite{attractive-mean-field-1,attractive-mean-field-2}.
Despite an early Bethe Ansatz solution~\cite{LL} for the continuum Bose gas 
with contact interactions, the exact treatment of attractive quantum systems lags 
behind the study of similar repulsive systems~\cite{Big-Hubbard-book,Mattis-red-book,Korepins-book}.

In this work we consider the one-dimensional periodic Bose-Hubbard model in the attractive regime, as a simple
boson model with short-range interactions and local hopping term. This model is in general not integrable~\cite{Haldane-Choy},
but it possesses several integrable limits and displays rich transitional behavior in the ground state.
A quantum phase transition (QPT) is usually defined as a phase transition at $T=0$ (i.e. in the ground state) 
under the variation of external
parameters, here the attractive interaction strength.
Phase transitions involve taking the thermodynamic limit, 
e.g. having infinitely many particles $N$ and lattice sites $L$.
Attractive boson systems are conceptually different
from repulsive bosons
and attractive/repulsive fermions, in that such a limit cannot easily be defined
as discussed later on.
Nevertheless, for large but finite 
$N$ and $L$ the attractive boson system does display an increasingly sharp distinction beween ground-state regions, 
similar to finite size realisations of system. 
We will within this paper denote this generalisation by {\sl pre-transition} and discuss its relevance as a tool
for the analysis of attractive bosons. 

To characterise the ground-state phases of the model, we study two key indicator properties.
The first is simply the energy gap between the ground 
and first-excited states. 
For finite systems the gap never vanishes, and there is never an occurrence of ground-state broken-symmetry in the quantum model as a result.
However we do observe through numerical analysis that the order of magnitude of the gap can be significantly different
across different coupling regimes, which leads to a sense of relative quasi-degeneracy~\cite{Kanamoto}.
The second key property we study is the incremental ground-state wave
function overlap, or the {\it fidelity} to use the language of quantum information theory.
% 
%
%  {
Recently there have a been a number of papers that have
used this concept to study quantum phase transitions in the thermodynamic limit~\cite{Zanardi,johnpaul}.
The essence of this approach lies in the fact
that if two states lie in different quantum phases then they are reliably distinguishable,
for example through the use of an order parameter. 
If states are
reliably distinguishable then they must be orthogonal~\cite{nielsen} and consequently the wave function overlap vanishes.
% 
% 
% }

For finite systems we propose
to modify this approach by identifying pre-transitions at couplings for which the incremental wavefunction overlap is (locally) minimal,
see \fref{fig:BH-1st-EV}.
For systems which exhibit a quantum phase transition in the thermodynamic limit it is then necessary that 
the value of the minimum goes to zero in the thermodynamic limit. 
In this manner we can say that the occurrence of a minimum in the incremental ground-state wavefunction overlap in a finite
system is a precursor for the quantum phase transition in the thermodynamic limit.
% 
% Technical issues in defining the incremental overlap for finite systems, like centering the difference, using a proportional
% or absolute difference between arguments as well as the usage of units in graphs, 
% are discussed in ~\cite{Semi-classical-method}.
Note that the incremental overlap graphs are shown on a unitless axis
% the actual values depend on the mentioned technical issues, but
as the physical interest here lies in the existence and location of minima, not the
quantitative shape~\cite{Semi-classical-method}.

The results we find from the study of incremental ground-state wavefunction overlaps give overwhelmimg support 
to the mean-field results, viz. the general existence of two transitional couplings. 
Within the context of mean-field theory the system exhibits a broken symmetry phase.
Our mean-field results point towards the existence of two transition couplings, with the critical couplings becoming degenerate at zero coupling 
in the limit of large particle number $N$ and a large number of lattice sites $L$. However by judiciously choosing     
the scaling of the parameters our findings also show that the limits
$N\rightarrow \infty$ and $L\rightarrow\infty$ do not commute. For example the bosonic statistics that underly the system mean that it is possible
to take $N$  to infinity while keeping $L$ finite. Moreover, we can also take $N$ and $L$ to infinity such that the ``density'' $N^{\mathcal D}/L\rightarrow$
constant for {\it any} ${\mathcal D}>0$. This prospect leads to the conclusion that the thermodyamic limit of the model appears to not be well-defined.
This is a significant distinguishing feature compared to fermionic lattice systems such as the Hubbard 
model~\cite{Big-Hubbard-book}
where the thermodynamic limit is well-defined.
In \sref{sec:two} we introduce the one-dimensional Bose-Hubbard model, list key properties used for this study
and present some numerical results for small systems.
Next a first analysis of the pre-transition points is given via different mean-field approaches in \sref{sec:MFT}, especially
the limiting case $L=2$ and $L\to\infty$ are discussed.
Section \ref{sec:Limiting_Integrable_Models} discusses the limiting solvable cases for $L=2$ (Bose-Hubbard dimer),
$N=2$ (Haldane-Choy),  and \mbox{$L\to\infty$} (Lieb-Liniger) via the Bethe Ansatz solution.
The results of the mean-field theory and the small size exact diagonalisation are compared with these exact solutions
and the limiting quasi root distribution is discussed.
The discussion in section \ref{sec:Discussion} finally puts all three approaches together and concludes
that in limiting cases, e.g. very small or very large $L$, only two regions might be visible.
Nevertheless, our main finding is that three ground-state phases exist in the attractive regime of model
for the generic case of finite but large number of particles $N$ and lattice sizes $L$ - presumably the 
experimentally relevant case.
% }
% 
\section{Definition of the Model}
\label{sec:two}
We consider a one-dimensional Bose-Hubbard model, consisting
of bosons with creation (annihilation operators)
$a_j$ ($a^\dagger_j$)
that create (annhiliate) a boson at lattice site $j$, with $j$ running over all $L$
lattice sites. The usual
bosonic commutation relations such as
$[a_j , a^\dagger_k]=\delta_{jk} I$ apply.
For a discussion of the physical origin and limitations of this model see~\cite{exp,BH-optical-lattice-realisations}.
Particles on the same site interact with interaction strength $\gamma$.
The kinetic term is given by nearest-neigbour hopping 
with coupling strength $t$, and 
periodic boundary conditions $a_{L+1}\equiv a_1$ are imposed.
% 
% \subsubsection{Real Space formulation}
In the real space presentation the Hamiltonian is given by
\begin{align}
H_\textrm{\tiny BH}=-t \sum_{j=1}^L 
\left[
a^\dagger_{j}a_{j+1}+a^\dagger_{j+1}a_{j}
\right]
% + \frac{\gamma}{N}
-\gamma
\sum_{j=1}^L
a^\dagger_{j}a^\dagger_{j}a_{j}a_{j}
% \nonumber
\label{eq:BA-Ham-real-space}
\end{align}
The Hamiltonian commutes with the total particle number $N=\sum_{j=1}^L n_j$ with $n_j=a_j^\dagger a_j$.  
The physical Hilbert space is spanned by Fock states of on-site occupation numbers
$|n_1,n_2,...,n_L\rangle$ with $a^\dagger_j |n_1,...n_j,...n_L\rangle=\sqrt{n_j+1}|n_1...n_j+1,...n_L\rangle$.
Its dimension ${d}=(N+L-1)!/((L-1)!N!)$ grows very rapidly with particle number and lattice size. 
For example, the moderate values $N=10$ and $L=20$  give the dimension of the Hilbert space as ${d}=20,030,010$,
strongly limiting exact diagonalisation of systems except for the dimer and trimer system.
Use of truncation schemes for the dimension and quantum Monte Carlo methods are limited by apriori unknown
behavior in the transitional regions of interest.

As the Hamiltonian (\ref{eq:BA-Ham-real-space})
conserves the momentum~\cite{Buonsante},
the matrix representation in a free momentum basis is
block diagonal,
%  in boxes with constant momentum. 
and low-lying, quasi-degenerate states are characterised by differing momenta.
Defining creation and annihilation operators in momentum space via the Fourier transforms
\mbox{
% 
% \begin{align}
$
b_k =
{{L}^{-1/2}}
\sum_{j=1}^L
\exp{(
\mathrm{2i}
 k j {\pi}/{L})
}
a_j$,
$ k=1...L$,
} it can be shown these operators satisfy canonical bosonic commutation relations
$[b_j , b^\dagger_k]=\delta_{jk}I$. The Hamiltonian (\ref{eq:BA-Ham-real-space}),
acting on a dual lattice of equally $L$ sites (modes)
may be equivalently expressed as
\begin{align}
H_\textrm{\tiny BH}=&\  -2t \sum_{k=1}^L 
\cos \!\left(\frac{2\pi k}{L}\right)\    b^\dagger_{k}b_{k}
\label{eq:BH-Ham-momentum}
\\
 \nonumber
&\qquad - \frac{\gamma}{L}
\sum_{k,l,m,n=1}^L
\!\!\!\!
b^\dagger_{k} b^\dagger_{l} b_{m} b_{n}
\delta_{k+l = m+n~(\textrm{mod} L)}
% \nonumber
% 
\end{align}
\begin{figure}
\includegraphics[angle=0,width=.94\linewidth]{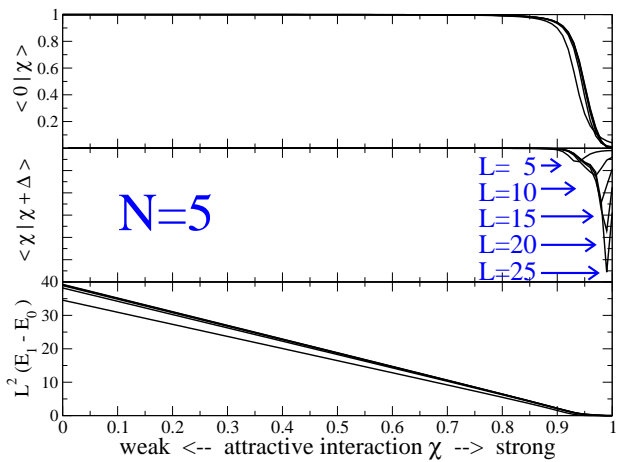}
\caption{
		Results of exact numerical diagonalisation of the Bose-Hubbard Hamiltonian (\ref{eq:BA-Ham-real-space})
		with $N=5$ bosons and $\epsilon_\chi=1$ for various numbers of lattice sites $L$ (order indicated by arrows holds for all panels).
		The properties shown are indicators of qualitative changes in the ground state 
		(cf.~\cite{Zanardi,johnpaul,Kanamoto}): (top to bottom) the ground-state overlap
		with the non-interacting ground-state {\footnotesize$|\chi\!=\!0\rangle$},\ 
		the incremental ground-state overlap {\footnotesize$\langle \chi| \chi+\Delta\rangle$}
		(for $\Delta=10^{-2}$), and the first excited energy relative to the ground-state energy.
		For explanations of the unitless axis in the middle graph see text.	
		In this particular parametrisation
	    	the transitional behaviour at $\chi_{c1}\approx 0.9$, predicted by the non-linear Schr\"odinger equation 
		approximation dicussed in Section \ref{subsec:NLS}, is visible. 		}
\label{fig:parametrisation-comparison}
\end{figure}
\begin {figure}[t!]
\begin{minipage}{\linewidth}
\begin{center}
\includegraphics[angle=0,width=0.9\linewidth]{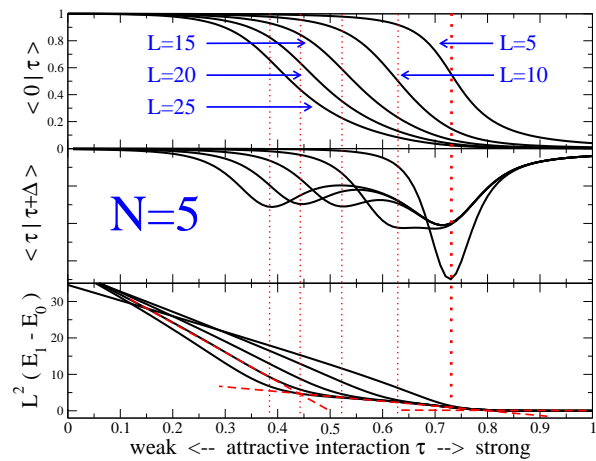}
\end{center}
\end{minipage}
 \caption{
The same data as in \fref{fig:parametrisation-comparison}, parametrised in terms of $\tau$.
The dotted lines are guide to the eye, to mark
three different regimes, the order of parameter $L$ indicated by arrows holds for all three panels.
In this parametrisation the $L$-dependence of the pre-transition coupling $\tau_{c1}$
is apparent, as indicated by the thin dotted 
vertical lines. 
The pre-transition coupling $\tau_{c2}\approx 0.73$, as indicated by the single thick dotted vertical line,
is independent of $L$ (cf.~\cite{Buonsante}). At
 this coupling we see a minimum of the incremental ground-state overlaps and the onset of quasi-degeneracy
 of the ground and first energy levels. 
The numerical value is not in close agreement with the predicted value $\tau_{c2}=2/3$ of \ref{subsec:SC}.
This discrepancy may be explained by the
fact that the particle number here is $N=5$, while the analysis in \sref{subsec:SC} assumes large particle number. 
}
\label{fig:BH-1st-EV}
\end{figure}

For the remainder of this paper we only consider $t>0$ and $\gamma>0$ corresponding to attractive interactions.
The model then incorporates the competition between the delocalising and localising
effects of the kinetic and the interaction terms respectively. 
In the limit \mbox{$\gamma\!\to\!0$} the ground state approaches that of non-interacting bosons, and is non-degenerate. 
At the other limit
\mbox{$t\!\to\!0$} the ground state becomes $L$-fold degenerate where the ground states consists of 
$N$ localised bosons on a single lattice site, viz. states of the form $|0,...0,n_j=N,0,...0\rangle$. 
However for non-zero $t$ the degeneracy is broken and the unique ground state is a superposition
of these localised states, giving rise to a Schr\"odinger cat state. The lowest $L$ energy states
 in this strong interaction limit form a narrow energy band. Within mean-field theory treatments,
 as will be shown below, this energy band degenerates
at non-zero values of $t$ giving rise to spontaneously broken translational invariance of the ground 
state. This provides the means to identify the ground-state phase boundaries. 
It had been realised~\cite{Kanamoto,Buonsante} that choosing 
interaction parameters depending on $N$ or $L$
keeps the regions of interest centered, see \fref{fig:parametrisation-comparison} and \fref{fig:BH-1st-EV}.
% % 
% 
% 
For the study of crossovers in the ground state the over-all energy scale can be neglected,
and we introduce parametrisations mapping the whole region from the weak to the strong coupling limit
into the finite interval $[0,1]$. As a dimensionless coupling parameter of the model we define $\delta=\gamma/t$, to study the
 ground-state properties of the model as $\delta$ is varied.
To help cope with the different scaling of the regions of interest as seen above, we introduce two further
 parametrisations in terms of dimensionless variables
$\chi,\,\tau\in[0,1]$. These are defined by 
\begin{eqnarray}
&&t=\epsilon_\chi(1-\chi),\qquad \qquad \gamma=\frac{\epsilon_\chi\chi}{NL}, \\
&&t=\epsilon_\tau(1-\tau),\qquad \qquad \gamma=\frac{\epsilon_\tau\tau}{N}
\end{eqnarray} 
where $\epsilon_\tau,\,\epsilon_\chi$ provide the energy scale. In terms of $\delta$ we have 
$$ \chi=\frac{NL\delta}{1+NL\delta}, \qquad \qquad \tau=\frac{N\delta}{1+N\delta}. $$
The non-interacting case is given by $\tau=\chi=0$ 
while pure interaction and no kinetic (hopping) contribution corresponds to $\tau=\chi=1$.
Other parametrisations, e.g. logarithmic dependence, are also used in the literature~\cite{Lundh}.
See \fref{fig:parametrisation-comparison}
and \fref{fig:BH-1st-EV} displaying the same information, for a visualisation of the effect of the parametrisation.
Numerical exploration for small systems finds that the dips (local minima in the incremental ground-state overlap) are quasi-stationary for
scalings of $\gamma\sim \frac{1}{N}$
and of $\gamma\sim \frac{1}{NL}$, respectively.
\section{Mean-Field Theory}
\label{sec:MFT}
Owing to the difficulties in treating the full quantum system with numerical and exact methods,
approaching the system in the spirit of mean-field theory has been very popular.
In particular with regards to investigating non-linear phenomena like solitons, and describing realistic experiments
on BECs,  these systems have been well studied in a wide range of contexts.
The continuum limit, known as the Gross-Pitaevskii equation~\cite{GPE-papers}, the Lieb-Linger Bose gas
or simply the non-linear Schr\"odinger equation (NLSE), have found wide interest~\cite{continuum-MFT-papers,Kanamoto}.
An extensive discussion of the mathematics of solution and further references 
for the {\sl discrete} NLSE
can be found in~\cite{discrete-NLS}.
For the purpose of this paper we are solely focused on pre-transitions in the ground state, though, and we will not
consider these applications here.

As the full discrete model is not integrable we will consider
the three cases of the dimer ($L=2$), trimer ($L=3$) and the continuum limit ($L\to\infty$).
We will then compare these special cases with  numerical solutions to the discrete mean-field equations for generic lattice size $L$.
In the last part of this section we will present a semi-classical analysis, following a different approach~\cite{Carruters}.
We will see that it recognises the second pre-transition not visible in the continuum limit,
at least qualitatively, i.e. the critical interaction scales correctly with $\sim\frac{1}{N}$,
compared with $\frac{1}{NL}$ in the continuum case.
\subsection{Generic $L$ and $N$ }
\label{sec:Generic-MFT}
Consider the Heisenberg equations of motion for the annihilation operators $a_j$
in (\ref{eq:BA-Ham-real-space})
\begin{align}
\mathrm{i} \frac{\textrm{d} a_j}{\textrm{d}t} = [a_j, H]\ , \qquad j=1...L
\nonumber
\end{align}
with restrictions to stationary solutions $a_j(t)=\exp({-\mathrm{i} E t})a_j$
for some energy eigenvalue $E$. We make the usual mean-field approximation,
here expressed by replacing the operators by complex numbers
\begin{align}
\nonumber
\vspace{-.5cm}
a_i,
a_i^\dagger
\qquad
&
\qquad
\textrm{\bf \huge$\longrightarrow$}
\qquad
&
a_i,a_i^* 
\qquad\qquad
\\[-.6cm]
\textrm{(operators)} 
&
&
\nonumber
\textrm{(complex numbers)} 
\end{align}
The resulting $N+1$ coupled equations
in the $N+1$ variables $a_1...a_N, E$
for the finite lattice are then given by
\begin{align}
% E\ a_1 =& (\tau-1)\, (a_{N} + a_{2}) &-& \frac{2 \tau}{N} |a_1|^2 a_1
% \nonumber
% \\
\nonumber
E\ a_j =& -t\, (a_{j-1} + a_{j+1}) - {2 \gamma} |a_j|^2 a_j\ , \quad j=1,...,L,
% \\
% E\ a_N =& (\tau-1)\, (a_{1} + a_{N-1}) &-& \frac{2 \tau}{N} |a_N|^2 a_N
% \nonumber
% \qquad
% i=1...N
\\
N=&\sum_{j=1}^L |a_j|^2 
% \nonumber
\label{eq:MFT-system}
\end{align}
Note that we will discuss this procedure in Section \ref{subsec:NLS} for the continuum model again. Either way,
taking the mean-field approximation first and then going to the continuum, or alternatively taking the continuum limit to the quantum
Lieb-Liniger gas and afterwards replace operators by complex numbers, the result is the same continuum Gross-Pitaevskii equation. 

\begin {figure}[tp]
% \hspace{-2cm}
\begin{minipage}{\linewidth}
\begin{center}
\includegraphics[angle=0,width=0.99\linewidth]{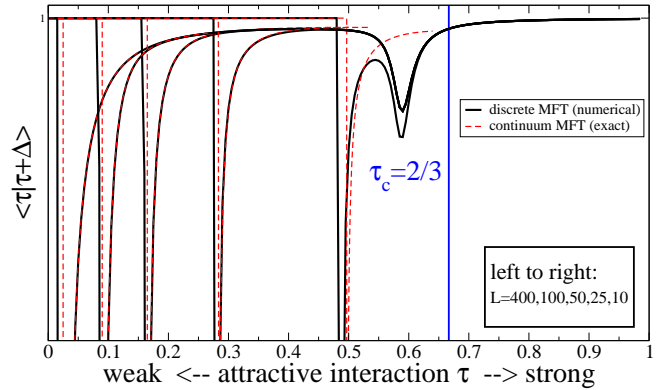}
\end{center}
\end{minipage}
 \caption{
 		Ground-state wave function overlap within the mean-field model of \sref{subsec:NLS}, versus
 		attractive interaction strengh $\tau$.
 		The numerical solution of the discrete non linear Schr\"odinger equation 
 		(solid black lines) exhibits the two dips locating the two transitional points,
 		already anticipated by \fref{fig:MFT-mom-dist} and \fref{fig:MFT-real-dist}.
 		The continuum approximation via the exact soliton solution 
		(\sref{subsec:NLS}) of the 
 		integrable non-linear Schr\"odinger equation (dashed red lines) 
 		has only one transitional point: this finite lattice effect is found in both 
 		the full quantum model and its semiclassical mean-field approximation when 
 		comparing with each respective continuum limit. 
 		Note that the exact mean-field theory result does drop off again for strong
 		interactions, which is not shown here.	
		Confer \fref{fig:LL-HC-incremental-overlap}
 		for the exact solution in the case $N=2$ (Haldane-Choy).
 		For strong interaction in this approximation the 
 		ground state does not enter a region of small changes again, thus not specifying
 		a second transitional point, nor does it relate to the location of the
 		lattice model transitional point, see graph for $L=10$.
 		}
 \label{fig:comparison-MFT-discrete-continuum}
\end{figure}

We show a numerical solution of these equations in \fref{fig:MFT-mom-dist}
 and \fref{fig:MFT-real-dist}.
Clearly the limiting case of (de)localisation in the densities can be seen, as well as one distinct and one less sharp crossover
within this mean-field theory. A clearer picture of the pre-transitions in the semi-classical analysis 
is given by the indicator property shown in \fref{fig:comparison-MFT-discrete-continuum}. The two dips 
scale the same, namely
$\sim\frac{1}{N}$ and $\sim\frac{1}{N L}$ respectively, found from exact diagonalisation.
As the discrete system (\ref{eq:MFT-system})
is non-integrable the general solutions are not known except for special cases.
In the limit of weak interactions the ground state of the orginal system is given by the
state with all particles in the zero momentum mode (an ``ideal BEC''), corresponding to the constant solution
% \begin{align}
$a_i = \sqrt{{N}/{L}} , \quad i=1...N 
$,
% \nonumber
% \end{align}
with energy $E= 2(\tau -1-{\tau}/{L})$ (For this and the following section we set $\epsilon_\tau=1$). This
delocalised wave function is a solution to the mean-field system for all interaction strengths, but 
for stronger interaction the ground state becomes a localised solution. Higher energy solutions can be constructed by considering extensions
$ a_j = \sqrt{{N}/{L}}\exp({2\pi j\phi}/L)$ or sawtooth like amplitudes. Here we are only interested in the lowest energy solution $\phi=0$.
Although the particle number $N$ enters the system of equations as a parameter, this mean-field approximation 
shows ``sharp'' pre-transitions between phases regardless of the value of $N$.
In the following
we will first discuss the special cases of the dimer ($L=2$) and the trimer ($L=3$),
before considering the case of large $L$. This non-linear system has many
solutions for a given parameter $\tau$. The numerical solutions shown were obtained
by starting from a known limiting case and then iterating via small changes in $\tau$. As will be seen this
procedure may lead to spontaneous ``hopping'' to another solution once the current one ceases to exist.
\label{sef:MFT-1}
\subsection{Dimer $N=2$}
\begin{figure}[t!]
	\begin{center}
		\begin{minipage}{1\linewidth}
		\includegraphics[angle=0,width=.99\linewidth]{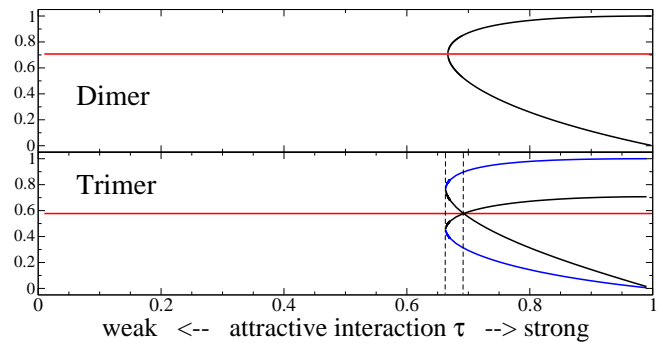}
		\end{minipage}
	\end{center}
\caption{
	Occupancy of the modes for varying attractive interaction $\tau$ within the discrete non-linear Schr\"odinger equation 
	approximation of section \ref{sef:MFT-1}.
	% 
	% 
% 	Motivated by looking for 
% 	specific solutions of the form 
% 	$a_1 > a_2 = a_3$ (``bright soliton'') and $a_1 < a_2 = a_3$ (``dark soliton''), 
%       the dimer ($L=2$) and trimer ($L=3$) can be treated analytically. 
        Here the upper (lower) line shows either $a_1=a_3$ ($a_2=a_3$).
	The horizontal line denotes the constant (``ideal BEC'')  solution, which exists
	for arbitrary $L$ and $\tau$. In the dimer there exists a pre-transition coupling $\tau_{c2}=2/3$, beyond which
	a second, increasingly localised solution exists. For the trimer case the pre-transition coupling 
	in a two-height scenario	is 
	$\tau_{c2}\approx 0.663$ (left dashed line).
	Note that the initially dark soliton (black lines) turns into a bright soliton before it merges 
	with the (lowest lying)
	bright soliton (blue lines), cf. \sref{subsec:DarkSoliton} for details. At $\tau_{c2}$ these real solutions cease to exist, but there is no smooth
	connection to the constant solution as in the dimer case, indicating the lattice effect,
	see text for further discussion, and cf. \fref{fig:MFT-real-dist}
	}
\label{fig:MFT-comparison-a}
\end{figure}

 For the dimer system the mean-field equations  (\ref{eq:MFT-system}) consist of three coupled, non-linear equations
for the complex variables $a_1$, $a_2$ and the energy $E$.
Assuming real solutions is equivalent to finding the roots of a 4th order polynomial. Two solutions are real for the whole
interaction range $0<\tau<1$, these lead to $a_1 = \pm a_2$, with ``$+$'' being the symmetric ground-state solution
(the horizontal line in \fref{fig:MFT-comparison-a}).
But for values of the coupling 
$\tau>\tau_{c2}$
there opens up two new solutions with (the same) lower energy. These connect at $\tau_{c2}$ to the constant solution 
$a_1= a_2$.
For $\tau\to1$ the solution localises, i.e. $a_1\to1$, $a_2\to 0$ or $a_1\to 0$, $a_2\to 1$,
as shown in \fref{fig:MFT-comparison-a}. 
The critical value $\tau_{c2}$ agrees with the semi-classical result of Sect. \ref{subsec:SC}
and an alternative mean-field treatment given in~\cite{Zoller}.

\begin{figure}[t!]
	\begin{center}
	\includegraphics[angle=0,width=.99\linewidth]{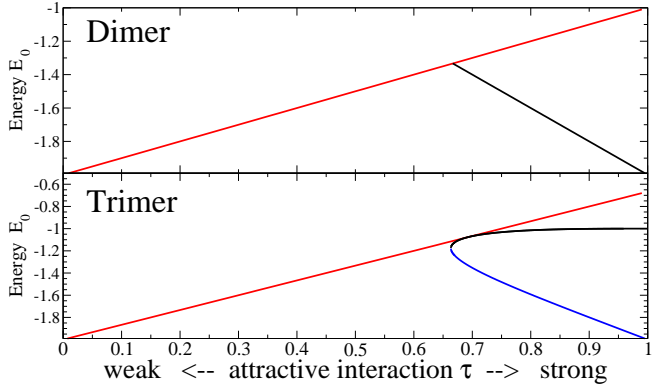}
	\end{center}
\caption{
	The energy for the lowest lying state of the two-height mean-field solution
	for dimer and trimer, corresponding to \fref{fig:MFT-comparison-a}.
	Note that for $L=2$ the soliton solution connects smoothly to the constant solution. 
	In the two-height approximation for the trimer there is already a small region around $\tau\approx0.663$
	where the true ground state is not of the two-height form. Compare this to the large middle region visible in
	\fref{fig:MFT-mom-dist} and \fref{fig:MFT-real-dist}, which differs from the conclusions in a recent study~\cite{Buonsante}.
	}
\label{fig:MFT-comparison-b}
\end{figure}
% 

% Note that the localisation does not get sharper or quicker for increasing particle number
% $M$, this is consistent with the observation in the (numerical) momentum distribution in the dimer!
% 
% 
% 
% 
% ############### TRIMER MFT     ########################################
% 
% 
\subsection{Trimer $L=3$}
\label{subsec:DarkSoliton}
The trimer system is non-integrable and it has been studied previously in the context of chaotic 
behaviour~\cite{Milburn,Franzosi}. Here we are only interested in soliton solutions
for the ground state within the mean-field description of the discrete non-linear Schr\"odinger equation.
It is useful to introduce the notion of bright and dark solitons. A bright soliton has a localisation
with positive amplitude relative to the constant solution, while the dark solition has a negative amplitude, i.e. a bright soliton looks like a hill and a dark soliton looks like a valley. For the dimer case the twice degenerate
bright soliton solution is at the same time a dark soliton, as the hill and valley cannot be distinguished
for $L=2$. For the trimer case $L=3$ bright and dark solitons have different energies.

Using a similar ansatz as for the dimer, i.e. $a_2=a_3$ and requiring
$a_i$ to be real, reduces the problem to the analysis of a 4th order polynomial.
A dark soliton ($|a_2|>|a_1|$) and a bright soliton ($|a_2|<|a_1|$) exist
for $\tau > \tau_a\approx0.663$, with the bright soliton having the lowest energy.
The energies of the two solitons are the same at $\tau = \tau_b\approx 0.692$, 
which is {\it not} the point at which the bright and dark 
degenerate into the constant solution, i.e. $|a_1|\neq|a_2|$ at $\tau = \tau_a$.
We remark further that at $\tau=\tau_b$ the dark soliton becomes a second, higher energy, bright soliton. 
At this point the energies for this soliton and the constant solution are the same, as shown in 
\fref{fig:MFT-comparison-b}.
This ceasing of the real solution of the form $a_1 > a_2 = a_3$ is a hint to the qualitative difference between the dimer
case $L=2$, and the general case $L>2$. We expect that in a small region the ground state is neither of the constant nor of the 
simple two-heights soliton form,
but a more complex solution connecting these both. 
This intermediate region does not exist for the dimer.

% 
% 
% % ############### CONTINUUM MFT     ########################################
% 
% 
% 
% 
% {
% 
% % 
% 
\subsection{$L\rightarrow\infty$: Non-linear Schr\"odinger equation approximation}
\label{subsec:NLS}
In the limit as $L\rightarrow\infty$ we can approximate the 
Bose-Hubbard Hamiltonian (\ref{eq:BA-Ham-real-space}) by a quantum field theory, with field operator $\Psi(x)$
satisfying
$$[\Psi(x),\Psi^\dagger(y)]=\delta(x-y).$$
Setting $a_j = \sqrt{\Delta}\Psi(j\!\cdot\! \Delta)$,
this consists of replacing $\Delta\sum^L \to \int^{\Delta L} \textrm{d}x$ under the assumption that $\Delta\ll 1$.
This is to be understood as choosing $L$ very large and $N$ finite, distinct from the usual notion of the thermodynamic limit where
$N, L\to\infty$ while keeping ${N}/{L}=$ constant.
The implication of this approximation will be discussed later. These considerations lead to a mapping of the Bose-Hubbard 
Hamiltonian to the non-linear Schr\"odinger equation, where the latter reads 
\begin{align}
H_{NLS}=& \int_{0}^{l} 
	%\hspace{-0.6cm}
	\left[
	\Psi'^\dagger(x)\Psi'(x)  
	% 
	% 
	% 
% 	+
	% 
	% 
	% 
	% 
	\right]
	\textrm{d}x
\nonumber
	\\[.25cm]
% 	\nonumber
	&\quad -
	c 
	\int_{0}^{l} 
	%\hspace{-0.6cm}
	% 
	\Psi^\dagger(x)\Psi^\dagger(x)\Psi(x)  
	\Psi(x)  
	\textrm{d}x
\label{eq:NLS-model}
\end{align}
with the periodic boundary condition $\Psi(0)=\Psi(l)$. 
At this point we remark that the Hamiltonian (\ref{eq:NLS-model}) is integrable \cite{Korepins-book} - 
see section \ref{sec:LL-BA} for more details. One of the conserved operators
is the total particle number
\begin{equation}
{\mathcal N}=\int_0^l \Psi^\dagger(x) \Psi(x) \textrm{d}x
\end{equation}  
which is quantised and has eigenvalues which are non-negative integers. The approximation of the Bose-Hubbard Hamiltonian by the non-linear Schr\"odinger equation is

\begin{eqnarray*}
H_{BH}&\approx& t\Delta^2 H_{NLS} -2t{\mathcal N}  \\
N&\approx&{\mathcal N}
\end{eqnarray*}
where $l=\Delta L$ and $c=  \gamma/\Delta t$. Hereafter we set $l=1$ or equivalently $\Delta=L^{-1}$.

The time evolution of the field operator $\Psi$ can be determined in the usual way :
\begin{eqnarray}
\mathrm{i}\frac{\partial \Psi}{\partial t} &=& [\Psi,H_\textrm{\tiny NLS}]
\nonumber
\\
&=& -\frac{\partial^2 \Psi}{\partial x^2} -2c|\Psi|^2\Psi \label{eq:time}
\end{eqnarray}
Our next step is to treat (\ref{eq:time}) as a classical field equation (cf.~\cite{Kanamoto}). We introduce 
the rescaled field \mbox{$\Phi=\sqrt{N^{-1}}\Psi$} and look for stationary solutions $\Phi(x,t)=\exp(-iEt)\Phi(x)$
such that 
\begin{align}
1 =& \int_{0}^1 |\Phi|^2 \textrm{d}x , 
\label{eq:class1}
% \end{align}
\\
% \begin{align}
E{\Phi} =& -\frac{\partial^2 \Phi}{\partial x^2} -U|\Phi|^2\Phi. 
\label{eq:class2}
\end{align}
where $U=2cN$.
The ground-state symmetry breaking solution to equations (\ref{eq:class1},\ref{eq:class2}) is known 
\cite{Kanamoto,attractive-mean-field-1}, and reads 
% 
%Apart from the constant, non-interacting BEC solution $\Phi=1$, which holds for all $\gamma$, 
%another exact solution~\cite{Kanamoto,attractive-mean-field-1} for strong interaction is of localised 'single-soliton' %form:
% 
% beyond a certain critical interaction strength $\tau_\textrm{\tiny c}$ there opens up a real-valued solution for $\Phi(x)$~\cite{Kanamoto,attractive-mean-field-1}.
% 
% 
% 
% 
\begin{align}
\Phi(x) \nonumber
\!=\!
\left\{
\begin{tabular}{ll}
$1$
 &
\ for $U<2\pi^2 $
\\
$\displaystyle \sqrt{\frac{K(m)}{E(m)}}\mathrm{dn}\!\left[ 2K(m)  (x\textendash x_0) | m\right]$
 &
\ for $U>2\pi^2 $
\end{tabular}
\right.
\end{align}
Above, $\mathrm{dn}(u|m)$ is a Jacobi elliptic function, $E(m)$ and $K(m)$ denote the 
complete elliptic integrals of the first and second kind, and $m$ is a function of 
$U$~\cite{Kanamoto,attractive-mean-field-1}.
Note that $x_0 \in [0,1]$ is the coordinate of the maximum of the wave function: the spontaneous symmetry breaking in the mean-field result
is visible from the degeneracy of the solutions beyond the point of ``collapse'' of the constant solution into a soliton at the critical coupling $U_c=2\pi^2$. In terms of the dimensionless coupling parameter $\chi$ of the Bose-Hubbard model this corresponds to 
\begin{equation}
\delta_{c1}=\frac{\pi^2}{NL}
\label{critical1}
\end{equation}
or equivalently
\begin{align}
\tau_{c1}= \frac{\pi^2}{L+\pi^2},\qquad\ \ 
 \chi_{c1}=\frac{\pi^2}{1+\pi^2}\approx 0.908
\end{align}
The soliton solution connects continuously, but not smoothly, to the constant solution at $\chi_c$ - 
in \fref{fig:comparison-MFT-discrete-continuum} this corresponds to dip in the dashed red line. A numerical solution 
for the finite size discrete NLSE
is shown \fref{fig:MFT-mom-dist} and \fref{fig:MFT-real-dist}, here this corresponds to the sharp change at
around $\tau\approx 0.2$.
% 
% which corresponds to
% % % 
% % 
% \begin{align}
% \tau_c =\frac{1}{1+\pi^{-2} }\approx -0.908
% \end{align}
% 
% In all presented mean field theory calculations the phase transition was between a region with a non-degenerate state
% and another region with
% $L$-fold degenerate ground state, constituting a spontaneous breaking of translational symmetry, which doesn't occur
% in the quantum model for finite particle number $N$.
% 
 % end green
% 
% 
% 
% 
% 
% {
\subsection{$N\rightarrow\infty$: Semi-classical analysis}
\label{subsec:SC}
In this section we present an alternative type of mean-field analysis, where we start by assuming that $N$ is arbitrarily large and $L$ is fixed.
 This is achieved by first canonically transforming to a number-phase representation of the quantum variables. 

Let $\{N_j,\theta_j\}_{j=1...L}$, 
obey canonical relations
$[\theta_i,\,\theta_j]=[N_i,\,N_j]=0$, \ $[N_j,\,\theta_k]=i\delta_{jk}$. We make a change of variables
% $b_j,\,b_j^\dagger,\,j=1,\,2$ via
\begin{align}
b_j=\exp(i\theta_j)\sqrt{N_j},
\quad 
b_j^\dagger=\sqrt{N_j}\exp(-i\theta_j).
\nonumber
\end{align}
Using the fact
$\exp(i\theta_j)N_j=(N_j+1)\exp(i\theta_j) $
it can be verified that the canonical commutation relations amongst the boson operators $b_j,\,b^\dagger_j$  are preserved. 
For large $N_j$ we can approximate the Bose-Hubbard Hamiltonian (\ref{eq:BA-Ham-real-space}) by
\begin{equation}
{\mathbb H}=-2t\sum_{j=1}^L \sqrt{N_j N_{j+1}}\cos(\theta_j-\theta_{j+1}) -\gamma\sum_{j=1}^L N_j^2
\label{classical}
\end{equation}
We now treat ${\mathbb H}$ as a classical Hamiltonian and look to minimise it subject to the particle number constraint
 \begin{equation}
 N=\sum_{j=1}^L N_j
 \label{eq:constraint}
 \end{equation}  
The minimum occurs when $\theta_j=\theta\,\forall j$, which leads us to studying 
\begin{align}
{\mathbb H}= {\mathbb H}_1 + {\mathbb H}_2 
\nonumber
\end{align}
where 
\begin{align}
{\mathbb H}_1&=&-2t\sum_{j=1}^L \sqrt{N_j N_{j+1}}\ ,
\qquad
{\mathbb H}_2&=& -\gamma\sum_{j=1}^L N_j^2
\nonumber
\end{align}
It can be verified that for $N_j=N/L\,\forall j$, ${\mathbb H}_1$ is globally minimal and ${\mathbb H}_2$ is globally maximal.
Thus for any $\gamma/t$ the solution $N_j=N/L\,\forall j$ provides a fixed point of the system, which will be the unique global
minimum when $\gamma/t$ is sufficiently small. We look to determine the coupling at which this solution ceases to be the minimum.
The results of the previous section indicate that when this happens a soliton solution will emerge. We can parametrise such a soliton solution as 
\begin{align}
N_j\geq \frac{N}{L} & \qquad {\rm for~} j \leq z, \\ 
N_j < \frac{N}{L} & \qquad {\rm for~} j>  z 
\end{align}
where $1\leq z \leq (L-1)$. Within this classical treatment, we can approximate the ground state for the full system by the two ground-state
configurations for the sublattices $j \leq z$ and $j>z$. For the full system at the pre-transition coupling $\delta_{c1}$ as predicted in \ref{subsec:NLS},
we see that the systems on the sublattices are below the pre-transition coupling due to the $L$-dependence of $\delta_{c1}$. 
Hence the ground-state configuration across each sublattice is one where the $N_j$ are constant on each sublattice.   This leads us to look
for soliton solutions within a {\it two-height} approximation, valid close to a 
point of broken symmetry:
\begin{align}
N_j=
\left\{
	\begin{tabular}{ll}
		$\displaystyle \frac{N(1+\alpha) }{2z} $ &  \qquad for  $j\leq z$ \\[.25cm] 
		$\displaystyle \frac{N(1-\alpha)}{2(L-z)}  $
		& \qquad for $j> z$ 
	\end{tabular} 
\right.
\nonumber
\end{align}
where $-1\leq\alpha\leq 1$ is continuous, such that 
(\ref{eq:constraint}) holds.

In terms of the above parametrisation the Hamiltonian is 
\begin{align}
{\mathbb H}=&
\nonumber
 -tN\left(\frac{(z-1)(1+\alpha)}{z} +\frac{(L-z-1)({1-\alpha})}{L-z}\right. 
\\
&\qquad\qquad\qquad + \left. 2\sqrt{\frac{1-\alpha^2}{z(L-z)}} \right) \\ 
% \\
&\qquad -
\frac{\gamma N^2}{4}\left( \frac{(1+\alpha)^2}{z}+\frac{(1-\alpha)^2}{L-z} \right)  
% s 
\nonumber
\end{align}
The next step is to minimise this expression for ${\mathbb H}$ with respect to the variables $z$ and $\alpha$, this involves mostly standard
calculus techniques.
We find that the smallest coupling for which symmetry-breaking of the ground state occurs is 
\begin{align}
\delta_{c2}= \frac{2}{N} 
\label{eq:MFT-crit},\qquad
\chi_{c2}= \frac{2L}{1+2L} 
% \nonumber
% \\
, \qquad
\tau_{c2}=\frac{2}{3}. 
% \nonumber
\end{align} 
In deriving the value $\delta_{c2}$, we justified the use 
of the two-height soliton approximation on the basis that $\delta_{c1}$ is a decreasing function of $L$. The fact that 
$\delta_{c2}$ was ultimately found to be independent of $L$ does not invalidate the two-height approximation within the context of the above analysis. 
The numerical results of \fref{fig:BH-1st-EV} illustrate that in a generic finite system the gap between the ground state and first excited-state
energy levels is {\it smaller} at $\delta_{c2}$ (or equivalently 
$\tau_{c2}$) than at $\delta_{c1}$ (or equivalently $\tau_{c1}$). Thus the notion of quasi-degeneracy of the energy levels is more appropriate at
$\delta_{c2}$ than at $\delta_{c1}$.  

Let us consider what happens when we now take the thermodynamic limit $N,\,L\rightarrow\infty$:  
\begin{align}
\delta_{c1}= 0,   & \qquad\chi_{c1}=\frac{\pi^2}{1+\pi^2}, & \tau_{c1}= 0,
\label{c1}  
\\
 \delta_{c2}= 0
,  &  
\qquad\chi_{c2}= 1, 
&  \tau_{c2}=\frac{2}{3} .
\label{c2}
\end{align}
The fact that the two sets of values (\ref{c1}) and (\ref{c2}) do not agree is an indication that the limits $N\rightarrow \infty$ and 
$L\rightarrow\infty$  do not commute, meaning that the usual concept of the thermodynamic limit is not well-defined
for this model. Equations (\ref{c1}) and (\ref{c2}) again show that two of the regions of interest 
will vanish when using the standard $\delta$ variables.
When using the parametrisation in the variables $\chi$ and $\tau$ only one region disappears
and one pre-transition point stays finite, i.e. away from 0 (no interaction limit) and 1 (no hopping limit).

Another curious point to observe is that while both $\chi_{c2}$ and $\tau_{c1}$ are $L$-dependent, 
they are in fact independent of $N$. 
This gives faith that the general qualitative ground-state features of the finite system will be tractable from 
analyses of systems with relatively small particle numbers. 
\section{Limiting Integrable Models}
\label{sec:Limiting_Integrable_Models}
\begin {figure}[t]
	\begin{minipage}{\linewidth}
		\begin{center}
		\includegraphics[angle=0,width=0.75\linewidth]{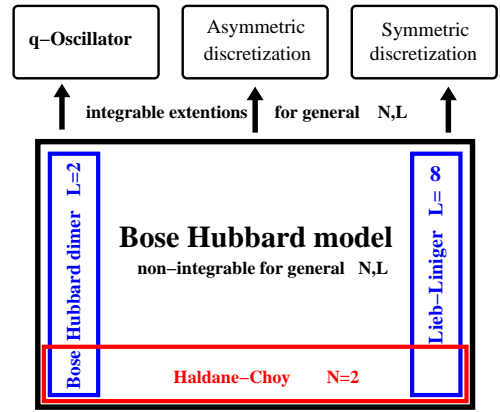}
		\end{center}
	\end{minipage}
	\caption{
	Integrable relatives of the Bose-Hubbard model: the 6 small boxes have Bethe Ansatz solutions, while the general
	Bose-Hubbard models is non-integrable.
 	}
\label{fig:integrable-extensions-scheme}
\end{figure}

% % 
% 
% 
% 
% 
While repulsive boson systems, as well as repulsive and attractive fermion systems
are well studied in the context of solvable systems the attractive boson gas
received comparatively little attention~\cite{attractive-integrable-LL-model}.
Still the seminal Bethe Ansatz solution
for one-dimensional contact interaction bosons~\cite{LL} in the continuum describes the repulsive as well as the attractive regime,
in which the solutions to the Bethe Ansatz equations become of different character ~\cite{McGuire-attractive-scattering}.
Initially it was believed that also the Bose-Hubbard model has a Bethe Ansatz solution~\cite{Haldane-Choy},
but it soon turned out that this model is non-integrable. Nevertheless there are several integrable limits and extensions,
see \fref{fig:integrable-extensions-scheme}.
Out of these we will examine the three limits shown inside the general 
 Bose-Hubbard box for the case of attractive interactions.
Integrable lattice distortions of bosons on a one-dimensional lattice, for instance the three boxes on top of  
\fref{fig:integrable-extensions-scheme},
have been studied mainly for the 
repulsive case~\cite{q-oscillator,Korepins-book,symmetric-discretisation,BH-quantum-chaos, Kundu}.
The attractive parameter region is technically harder
than the repulsive case: for instance the attractive Bethe Ansatz roots in the Lieb-Liniger model lack several of the
properties which allowed analysis for repulsive interaction: e.g. string solutions which keep their string form 
and saturation of the root distribution in the thermodynamical limit.
The problem of {\sl collapse} of the system already at infinitely weak attractive interaction when taking the thermodynamic limit
 is less problematic~\cite{Mattis-red-book} and has been addressed by the $N$-dependent reparametrisation~\cite{Kanamoto,Buonsante}.
% used in this study.
% 
% 
% 
In the next secion \ref{sec:BH-dimer} we will briefly discuss the Bethe ansatz solution of the $L=2$ dimer case and point out that
it has only one crossover. Then we move on to discuss the finite lattice size case $L>2$ via the two-particle solution
of the Haldane-Choy ansatz: here we see now two pre-transitions, i.e. dips in the indicator property ground-state overlap,
which have the correct scaling behaviour found numerically and via semi-classical analysis earlier in this paper.

We will establish via the $N=2$ Haldane-Choy solution and the results from \sref{sec:MFT} for the relation between
the (discrete) NLSE and the (continuum) GPE that the integrable Lieb-Liniger continuum gas is a good proxy for the discrete 
lattice model for the study of the $NL$-dependent pre-transition.
This motivates our discussion of the Bethe ansatz root distribution for the attractive ground state of the Lieb-Liniger model
and relates these results to the Bose-Hubbard model in the last section.
% 
% 
% 
% 
% 
% 
% 
% 
% 
% 
% 
% 
% {
\subsection{Bose-Hubbard dimer}
\label{sec:BH-dimer}
For the dimer case, $L=2$, the Hamiltonian (\ref{eq:BA-Ham-real-space}) reduces to 
\begin{align}
H=\textendash 2 t
\!
\left( a_1^\dagger a_{2}+a_{2}^\dagger a_{1} \right)
- \gamma
\left(
a_1^\dagger a_1^\dagger a_{1} a_{1} 
+
a_2^\dagger a_2^\dagger a_{2} a_{2} 
\right). 
\nonumber
% \label{eq:bhd}
\end{align}
This Hamiltonian can be expressed~\cite{enolskii1} in terms the $su(2)$ algebra with generators $\{S^z,\,S^\pm\}$ and relations 
$$ [S^z, S^\pm]=\pm S^\pm,\qquad [S^+,S^-]=2S^z.$$ 
Using the Jordan-Schwinger representation
$$S^+= a^\dagger_1 a_2, \quad
S^-= a^\dagger_2 a_1, \quad
S^z=\frac{1}{2}(a^\dagger_1 a_1-a^\dagger_2 a_2) $$ 
this leads to 
\begin{align}
H=- 2 t\left( S^++S^- \right)
- {\gamma}
\left(2(S^z)^2+\frac{1}{2}N^2-N
\right). \nonumber
\end{align}
The same $(N+1)$-dimensional representation of $su(2)$ is given by the mapping to differential operators
$$ S^z= u\frac{{\rm d}}{{\rm d}u}-\frac{N}{2},\quad  
S^+=Nu-u^2\frac{{\rm d}}{{\rm d}u}, \quad
S^-=\frac{{\rm d}}{{\rm d}u} $$ 
acting on the space of polynomials with basis $\{1,u,u^2,...,u^N\}$.  
We can then equivalently represent the dimer Hamiltonian $H$ as the second-order differential operator
\begin{eqnarray}
H &=&-2t\left(Nu+(1-u^2)\frac{{\rm d}}{{\rm d}u}\right) \nonumber \\ 
&&\quad -\gamma(2u^2\frac{{\rm d}^2}{{\rm d}u^2}+2(1-N)u\frac{{\rm d}}{{\rm d}u}+N^2-N) \nonumber \\
&=&-2\gamma u^2 \frac{{\rm d}^2}{{\rm d}u^2} 
+\left(2\gamma(N-1)u+2t(u^2-1)\right)  \frac{{\rm d}}{{\rm d}u}   \nonumber \\ 
&&\quad\qquad +\left(N\gamma-N^2\gamma-2tNu \right).
\label{bhde}
\end{eqnarray}
Now we look for solutions of the eigenvalue equation 
\begin{equation}
HQ=EQ
\label{eq:eigen}
\end{equation}   
where $Q$ is a polynomial function of order $N$ which we express in terms of its roots $\{v_j\}$:
$$Q(u)=\prod_{j=1}^N(u-v_j). $$
Evaluating (\ref{eq:eigen}) at $u=v_k$ for each $k$ leads to the set of Bethe ansatz equations 
\begin{align}
\frac{t(1-v_k^2)+\gamma(1-N)v_k}{\gamma v^2_k}=\sum^N_{j\neq k}\frac{2}{v_j-v_k},\,\, k=1...N.
\label{eq:bhd_bae}
\end{align}
By considering the terms of order $N$ in (\ref{eq:eigen}) the energy eigenvalues are found to be 
\begin{eqnarray}
E= \gamma N(1-N)+2t\sum_{j=1}^N v_j.
\label{eq:bhd_nrg}
\end{eqnarray} 
We can transform the differential operator (\ref{bhde}) into a Schr\"odinger operator~\cite{dhl}. Setting 
% \begin{eqnarray*}
\begin{align}
\Psi&=\exp\left(-\frac{t}{\gamma} \cosh(\sqrt{2\gamma}x)-N\sqrt{\frac{\gamma}{2}}x\right) 
\nonumber
\\
\nonumber
&\qquad\quad \times \prod_{j=1}^N(\exp(\sqrt{2\gamma}x)-v_j), 
\end{align} 
\begin{align}
\widetilde{H}=-\frac{{\rm d}^2}{{\rm d}x^2} + V(x), 
\nonumber
\end{align} 
where the potential $V(x)$ is 
$$V(x)= \left(\frac{2t^2}{\gamma}\sinh^2(\sqrt{2\gamma}x)-2t(N+1)\cosh(\sqrt{2\gamma}x)\right), $$
then 
$$\widetilde{H}\Psi= E \Psi $$ 
with $E$ given by (\ref{eq:bhd_nrg}) whenever the $\{v_j\}$ are solutions of the Bethe Ansatz equations~(\ref{eq:bhd_bae}). 
It is easily checked that the potential has a single minimum when $\gamma/t=\delta< 2/(N+1)$ and two minima when $\delta>2/(N+1)$. The critical value 
$\delta=2/(N+1)$ agrees, to leading order in $N$, with the mean-field theory result for $\delta_{c2}$ of section \ref{subsec:SC},
cf. equations~(\ref{eq:MFT-crit}).

\begin{figure}[tp!]
	\includegraphics[angle=0,width=0.99\linewidth]{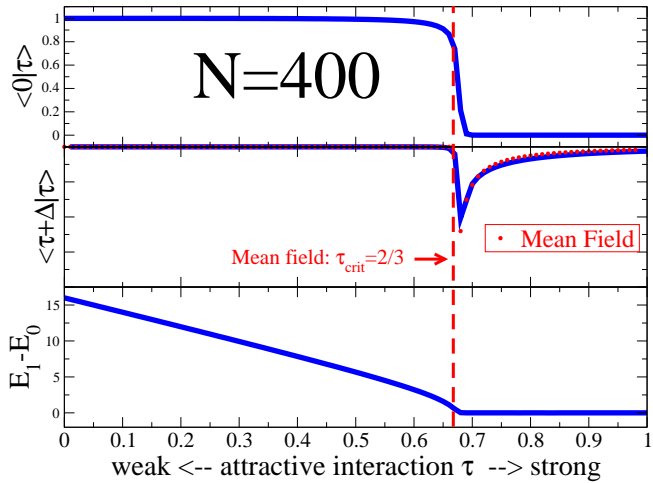}
	\caption{
			The Bose-Hubbard Dimer $L=2$ has a single pre-transition point. Shown are (top to bottom), 
			ground-state wave function overlap with the non-interacting reference state,
			the incremental  ground-state wavefunction overlap and the first excitation energy relative to the 
			ground-state energy, each
			obtained by exact numerical diagonalisation. The dashed line indicates the theoretical value of $\tau_{{\rm crit}}=2/3$ 
			which is given by mean-field theory.
			For the wavefunction overlaps in the middle graphic the mean-field result for the overlap are shown by the dotted line, 	
			which is barely distinguishable from the numerical result. The value obtained from  the exact Bethe ansatz solution is  
			$\tau_{BA}=2/(3(1+N^{-1}))$, giving the quantum correction to the mean-field result.  
			}
	\label{fig:BH-dimer-comparison}
\end{figure}

From the analysis of the Bethe Ansatz solution it can be seen the limiting case of only two lattice sites
has a single transitional point, visualised in \fref{fig:BH-dimer-comparison}.
This corresponds to the case where the two minima in \fref{fig:BH-1st-EV}
coincide and the middle region is no longer visible.
In \fref{fig:BH-dimer-comparison} two physically very different regimes can be seen:
the ground-state overlap measures the relative weight the occupation of the zero momentum
mode by all particles has (relative to the non-interaction BEC state with 100\% condensation at $\tau=0$). 
For small $\tau$ these contributions dominate, while after a small crossover region
for large $\tau$ this non-interacting BEC state has a very low relative weight in the ground state.
In the complementary plot, against the $N$-body Sch\"odinger cat-state $|N,0\rangle+|0,N\rangle$ as reference state,
the overlap would be almost constant, close to $1$ in the strong interacting regime on the right, while it would be 
$\ll 1$  and quasi-constant in the weak interacting region: the relative weight of a localisation of N particles
is low. Similar for the bottom picture in \fref{fig:BH-dimer-comparison}: the ground-state energy for the 
very weakly interacting regime $\tau\approx 0$ is non-degenerate, the first excitation is separated by the energy required to
transfer a single boson from the zero momentum mode to the first momentum. In the strong interacting limit for large
$\tau$ the ground state is quasi degenerate :  the (anti-)symmetric cat states $|N,0\rangle\pm|0,N\rangle$ have the same
energy.
The mean-field calculation for the dimer, see~\fref{fig:MFT-comparison-a}, shows the 
(square root of) the
relative occupancy of the two sites.
For $\tau$ below the critical interaction both sites are equally occupied, the totally delocalised constant solution, with
all the particle in the lowest momentum mode $b_0$. At $\tau_c=\frac23$ the symmetry breaks and one site has higher occpuation than the others.
Due to the quantum-mechanical superposition in eigenstates this can only be seen in the mean-field theory. The second momentum mode
has a finite and increasing occupation beyond the critical interaction, though, and it 
reaches $n_{k=0}=n_{k=1}=\frac12$ for $\tau\to1$. This is the complete delocalisation in momentum space and corresponds
to the complete localisation in real space density observed in the soliton solution.
\subsection{Haldane-Choy Bethe Ansatz for $N=2$}
% 
% \label{sub-sec:HC-BA}
\label{sec:HC-BA}
% \subsubsection{Bethe Ansatz solution}
The Bose-Hubbard model (\ref{eq:BA-Ham-real-space})
has a Bethe Ansatz solution in the spirit of the fermionic Hubbard model,
but it is only solvable for a maximum site occupation of two particles~\cite{Haldane-Choy}. For $N=2$
the exact eigenstates
%$|\textrm{BA}\rangle$
are
% given by
\begin{align}
\left| \mathrm{BA}\right\rangle =&
\sum_{i,j=1}^L
C_{ij}
|i,j\rangle
\nonumber
% \end{align}
% \begin{align}
\\
C_{nm}=&
\left\{
\begin{tabular}{ll}
$
\mathrm{e}^{
\mathrm{i} \left(
k  n + q  m
\right) 
}
+
\frac{
\sin k  - \sin q 
-\mathrm{i}
 \gamma
 }{
\sin k  - \sin q 
+\mathrm{i}
 \gamma
}
\mathrm{e}^{
\mathrm{i} \left(
q  n + k  m
\right) 
}
$
&
$
,\ \ 
n\leq m
$
\\
$C_{mn}$ 
&
$
,\ \ 
n>m$
\end{tabular}
\right.
\nonumber
% \end{align}
% \begin{align}
\nonumber
\end{align}
with the Bethe Ansatz equations
\begin {figure}[t!]
% \hspace{-2cm}
	\begin{minipage}{\linewidth}
		\begin{center}
		\includegraphics[angle=0,width=0.99\linewidth]{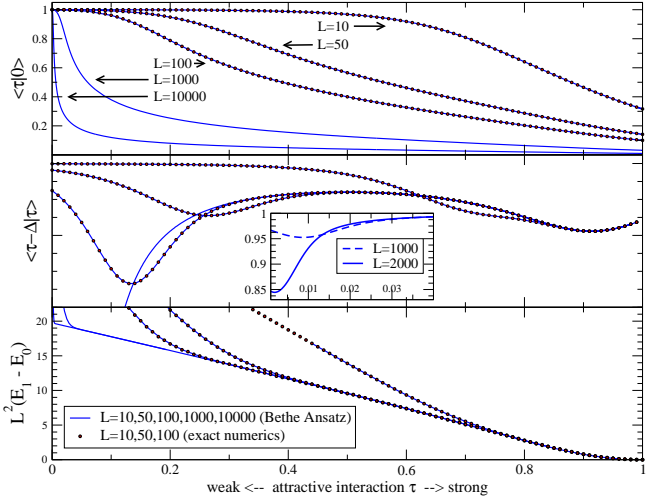}
		\end{center}
	\end{minipage}
 	\caption{
		Similar to \fref{fig:BH-1st-EV} for the solvable case
		of two bosons in a lattice of differing size $L$ (order indicated by arrows holds for all three panels) 
		Already for the minimal particle number $N=2$ the two pre-transition points can be clearly seen,
		compared to the one pre-transition point for the dimer, see \fref{fig:BH-dimer-comparison}. 
		The exact Bethe Ansatz solution (\ref{eq:BH-overlap}) (solid lines),
		available for arbitrary  $L$,  is compared with exact numerics (dots) for small sizes.
		%	 		.
		Shown are (top to bottom):
		the ground-state overlap with the non-interacting ground state {\footnotesize $|\tau\!=\!0\rangle$},
		the incremental overlap {\footnotesize $\langle\tau|\tau+\Delta\rangle$} for $\Delta=10^{-2}$,
		and the first excited energy relative to the ground state $L^2 (E_1 - E_0)$.
		%. 
		% 
		}
\label{fig:BH-HC-incremental-overlap}
\end{figure}
\begin{align}
\mathrm{e}^{\mathrm{i} k L}
=\frac{
\sin k  \!-\! \sin q 
\!-\!\mathrm{i}
 \gamma
 }{
\sin k  \!-\! \sin q 
\!+\!\mathrm{i}
 \gamma
}
\ \, 
\wedge
\ \,
\mathrm{e}^{\mathrm{i} q L}
=\frac{
\sin q  \!-\! \sin k 
\!-\!\mathrm{i}
 \gamma
 }{
\sin q  \!-\! \sin k 
\!+\!\mathrm{i}
 \gamma
}
% \nonumber
\label{eq:HC-BAE}
\end{align}
{
Here 
\mbox{
\!	$|i,j\rangle= 
% 2^{-\frac{\delta_{ij}}{2}}\cdot 
a^\dagger_i a^\dagger_j|0\rangle$} is not normalised, 
i.e. $\langle i,j|i,j\rangle= 1$ respectively $=2$ for $i\neq j$ respectively $i=j$.
}
The energy eigenvalues for these are $E=-2 (\cos k + \cos q)$, motivating the name ``quasi-momenta'' for the Bethe Ansatz roots $k$ and $q$.
The Bethe Ansatz roots for the ground state are symmetric,  $k=-q$, and imaginary for 
attractive interactions $\gamma>0$.
For this setting we can define $k=-q=\mathrm{i} K$, with $K>0$ determined by the single Bethe Ansatz equation,
here in inverse form
% 
% 	\begin{align}
% 	\qquad
% 	\mathrm{e}^{-K L} =
%  	\frac{2 \sinh K +  \gamma }{2 \sinh K - \gamma}
% % 	\nonumber	
% \label{eq:Single-HC-BAE}
% 	\end{align}
% 
% For this special case the inverse relation $K(\gamma)$ is known in closed form:
\begin{align}
\gamma(K)
=
2\sinh K \ \tanh \frac{K L}{2}.
\label{eq:inverse-HC-BA}
%  \qquad \gamma,K >0
\end{align}
For use in the next section we also note that for $\gamma>\frac{4}{L}\cos\frac{\pi}{L}$ the two real roots
of the BAE solution for the first excitation $E_1$ merge to a complex 2-string of the form
\mbox{$k,q=\frac{\pi}{L}\pm\mathrm{i} K$} with $K>0$, and the inverse function
given by \mbox{$\gamma(K)=2 \cos \frac{\pi}{L} \coth \frac{K L}{2}\sinh K$}.
The real roots of the first excitation for  $\gamma<\frac{4}{L}\cos\frac{\pi}{L} $ are given by
$k=K$, $q=\frac{2\pi}{L}-K$ with $0< K <\frac{\pi}{L}$. The inverse function relating the parameter $K$
to the interaction strength is 
$\gamma(K)=2 \cos \frac{\pi}{L}  \tan \frac{K L}{2}\sin \frac{K L -\pi}{L}$.
We use these expressions for the analysis of the indicator properties like $L^2 (E_1 -E_0)$,
in \fref{fig:BH-HC-incremental-overlap}, as well as for comparison with the
Lieb-Liniger continuum model in the next section.

The (not normalised) ground-state wave function can be written as, 
cf. (\ref{eq:LL-wavefunction}):
\begin{align}
|K\rangle=&
\sum_{n,m}
\left[
 \mathrm{e}^{K (\frac{L}{2}- |n-m| )}+\mathrm{e}^{-K (\frac{L}{2}- |n-m| )}
\right] |n,m\rangle
\label{eq:HC-wavefunction}
\\
\nonumber
=&
\sum_{n,m}
\left[
2 \cosh K ( |n-m| - \frac{L}{2})
\right] |n,m\rangle
\end{align}
resulting in the closed form expression for the (not normalised) overlap in the Haldane-Choy model
% {
% 
{
\begin{align}
% \nonumber
\label{eq:BH-overlap}
\left\langle
K\!+\!\Delta
|
K\!-\!\Delta
\right\rangle
=\hspace{4.6cm}
\\
\nonumber
% 
% 
% 
% \
% \frac{L}{2}\
%  % 
% % \mathrm{e}^{-L \frac{K+Q}{2} }
%  \cosh {K L}
%  \cosh {\Delta L}
%  \\
% \nonumber
% +&
% %  \mathrm{e}^{-L\frac{K + Q}{2}}
% \sum_{j=1}^{L \textendash 1}
% (L \textendash j)
% \cosh [K( L\textendash 2 j)]
% \cosh [\Delta(L \textendash 2 j)]
% % \right]
4 L \left( 
\coth K \coth K\,L
+
\coth \Delta \coth \Delta\,L
\right)
\end{align}
}
{
together with the normalisation
\begin{align}
\langle K|K \rangle
\nonumber
=
\nonumber
% \langle K | K \rangle
% =
 L \mathrm{e}^{-K L}
\left(
\coth K -1
\right)
% \times\cdots
\hspace{3cm}
\\
\times
\left(
2 L
\mathrm{e}^{K(L+2)}
-
\nonumber
2 L 
\mathrm{e}^{K L}
+
\mathrm{e}^{2 K (L+1)}
+
\mathrm{e}^{2 K L}
-
\mathrm{e}^{2 K}
-1
\right)
\end{align}
this results in the normalised overlap expression
\begin{align}
\frac{\langle K+\Delta | K-\Delta \rangle}{\sqrt{\langle K+\Delta | K+\Delta \rangle \langle K-\Delta | K-\Delta  \rangle}}.
\nonumber
\end{align}
}
In the above equations $K\pm\Delta$ denote the imaginary parts of  the single Bethe Ansatz root associated 
with the two different interaction strengths $\tau_1 \mapsto K+\Delta$ and $\tau_2 \mapsto K-\Delta$,
i.e. solutions to (\ref{eq:inverse-HC-BA}).
This expression depends on the interaction strength $\gamma$ only through the Bethe Ansatz root
$K$, allowing closed form solution in parametrised form~\footnote{
The relations between $K$, $\gamma$ and $\tau$ are non-linear - shown in the figures
are $\langle \tau+\Delta|\tau\rangle$, not $\langle K+\Delta|K\rangle$, the formula is given in terms of the Bethe Ansatz root K
and in symmetrized form to simplify the expression.
}.
The Bethe Ansatz solution for only two particles is not truly a many-particle solution - the $N=2$ Bose-Hubbard model 
can be treated exactly conventionally in center-of-mass coordinates~\cite{Fluegge,rep-BH-experiment}. 
In that case the physical meaning of the Bethe Ansatz quasi-momenta is lost, though. The solution presented here is visualised in 
\fref{fig:BH-HC-incremental-overlap}.
We remark that within this approach the exact momentum distribution of two bosons
in the one-dimensional lattice can be calculated explicitly, clarifying the connection between the (here two) Bethe Ansatz
quasi-momenta and the physical momenta, which is of interest for example in the integrable boson-fermion mixture~\cite{our-mixture}. 
% see also section \ref{sec:LL-BA-TBA}.
% 
% 
% 
% 
% 
% 
% 
% 
% 
% 
{
\subsection{Lieb-Liniger approximation}
\label{sec:LL-BA}
The continuum model in (\ref{eq:NLS-model}) is the integrable Lieb-Liniger gas~\cite{LL}.
For the repulsive regime it is arguably one of the best studied integrable 
models~\cite{LL,Mattis-red-book,Takahashi-book,Korepins-book,weak-bosons,LL-repulsive-studies},
while the attractive regime is less popular~\cite{attractive-integrable-LL-model}, due to
difficulties in taking the thermodynamic limit.
When taking the limit $L\to\infty$ in the Bose-Hubbard model the Lieb-Liniger
model can be used as an integrable approximation for the weak coupling limit. 
Information is lost in taking this limit, i.e. in going
from the three independent parameters $N,L,\gamma$ of the Bose-Hubbard model to the two parameter
continuum model. Thus we expect the range of validity to be restricted, but in turn the property of integrability is gained.
There are two different ways of looking at this integrable model as a limit of the Bose-Hubbard model,
we consider the analysis for finite $N$ and large $L\to\infty$ in section \ref{sec:LL-BA-finite-N}.
For {\sl repulsive} interactions the thermodynamic limit can very succesfully be treated (see~\cite{Korepins-book} for references)
for constant density $n=\frac{N}{L}$ when $N,L\to\infty$.
In the first case the two independent parameters are $N$ and an interaction strength. In the second case we keep
the particle density/filling factor $n$ and an interaction strength as free parameters.
This physical notion of density (of Bethe ansatz roots) is not extendable to the attractive case, the 
bosons tend to cluster up instead of saturating. We discuss this further in~\sref{sec:LL-BA-TBA}.	
\subsubsection{Analysis for finite $N$ and large $L\to\infty$}
\label{sec:LL-BA-finite-N}
In this section we analyze the special case $N=2$ as example for finite $N$ and (very much) larger $L\gg N$.
The exact Haldane-Choy solution discussed in section {\ref{sec:HC-BA}}
is the yardstick to explore the impact of the continuum approximation in the full quantum model. 
The energy eigenvalues in the general Lieb-Liniger model corresponding to an approximated lattice model  are given by
\begin{align}
L^2 E= 
% L^{-2}
  \sum_{i=1}^N k_i^2  
\
+
\textrm{const.}
\end{align}
with the $N$ complex parameters $k_i$ determined by solutions to the Bethe ansatz equations
\begin{align}
&\mathrm{e}^{\mathrm{i} k_i }
=&
% \nonumber
\prod_{j\neq i}^{N}
\frac{k_i - k_j - \mathrm{i} \tau\frac{L}{N}}{k_i - k_j + \mathrm{i}\tau\frac{L}{N}}\ ,
\qquad
i=1...N
 \label{eq:LL-BAE}
\end{align}
In particular we see that the continuum model gets mapped onto the weak coupling limit
of the lattice model, as $\frac{N}{L}\to 0$ for $N=2$ and $L\to\infty$. To check how well the 
Lieb-Liniger model approximates the Bose-Hubbard model for large $L$
we calculate analytically the ground-state overlap for the case $N=2$ and compare with the Haldane-Choy expression (\ref{eq:BH-overlap}).
The root behavior for ground state and first excitation (see also appendix of~\cite{LL}) is similar to the lattice case:
the two ground-state roots form again a purely imaginary complex pair $k_{1,2}=\pm\mathrm{i} K$, where the inverse function
is given by 
\begin{align}
\gamma = 2\, \frac{K}{L}\tanh \frac{K}{2}.
\end{align}
The first excitation roots form a complex pair past the interaction strength $\gamma > \gamma_c =\frac4L$ of the form 
 $k_{1,2}=\pi \pm\mathrm{i} K$, with inverse function $\gamma = 2\, \frac{K}{L}\coth\frac{K}{2}$.
For  weak interaction $\gamma <\gamma_c$ the first excitation has two real roots at 
$k_2=2\pi \textendash k_1=2\pi \textendash K$, with
inverse function $\gamma=\frac{2}{L}(\pi\textendash K)\tan\frac{K}{2}$.

The (not normalised) ground-state wave function for finite interaction and $N=2$ is given
by, cf. (\ref{eq:HC-wavefunction})
{
\begin{align}
\left|K\right> =&
\mathrm{e}^{+K (|x-y|-\frac12)}
+
\mathrm{e}^{-K (|x-y|-\frac12)}
\label{eq:LL-wavefunction}
\\
=&2 \cosh K(|x-y|-\frac12)
\nonumber
\end{align}
}
The {\it normalised} ground-state wavefunction overlap 
for two different interaction strengths, with corresponding imaginary part of roots  $K\pm\Delta$,
is given by, cf. (\ref{eq:BH-overlap}) 
\begin{align} 
&
% \int
 \left<K+\Delta|K-\Delta\right> 
% dx dy 
\label{eq:LL-overlap}
\\
\nonumber
=&\left( \frac{\sinh(K)}{K}+\frac{\sinh(\Delta)}{\Delta}\right) 
\\
\times&
\sqrt{\frac{K^2-\Delta^2}{(K-\Delta+\sinh(K-\Delta))(K+\Delta+\sinh(K+\Delta))}}
\nonumber
\end{align}
% integral in half.
The comparison for $N=2$ of the rescaled Lieb-Liniger gas with the Bose-Hubbard system is
shown in \fref{fig:LL-HC-incremental-overlap}.
\begin {figure}[t!]
	\begin{minipage}{\linewidth}
		\begin{center}
		\includegraphics[angle=0,width=0.99\linewidth]{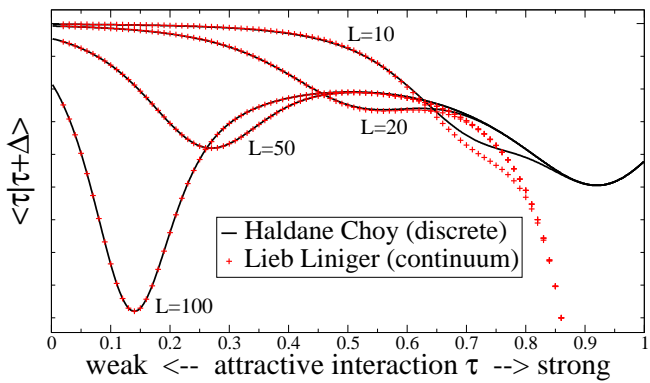}
		\end{center}
	\end{minipage}
	\caption{
		Ground-state wave function overlap versus 
		attractive interaction strength $\tau$ for $N=2$ bosons.
		The exact solution of the full quantum model (black solid lines)
		on the finite lattice (Haldane-Choy) exhibits two minima, 
		indicating two transitional points.
		The continuum approximation via
		the Lieb-Liniger model (red crosses) discussed in~\sref{sec:LL-BA-finite-N}
		displays only one minimum,
		indicating it has only one transitional point, see text and cf. 
		\fref{fig:comparison-MFT-discrete-continuum}
		for the large $N$ mean-field result.
		Note that the agreement is best in the $L$-dependent weak regime (left side) and not in the
		the $L$-independent strong interaction regime (right side). 
		}
	\label{fig:LL-HC-incremental-overlap}
\end{figure}
From the exact diagonalisation of small systems, see 
\fref{fig:BH-1st-EV} and \fref{fig:BH-HC-incremental-overlap},
it is apparent that for increasing lattice size $L$ and fixed particle
number $N$ a growing region extends from the strong interaction limit $\tau=1$ to smaller $\tau$.
The shown physical properties in this region are independent of $L$, nevertheless the 
$L\to \infty$ Lieb-Liniger model is not a valid approximation for that region:
from the Bethe Ansatz equations (\ref{eq:HC-BAE}) and (\ref {eq:LL-BAE})
it can be seen that the interaction strength gets rescaled 
by $\sim L^{-1}$, effectively mapping the Lieb-Liniger model onto the infinitely weak interacting 
Bose-Hubbard model by quasi-linearising the roots.
The continuum limit does not capture the physics in the strong interaction region, in particular it 
does not see the {\sl second} pre-transition point $\tau_{c2}$ connected to the finite lattice effect.

Our results for the semi-classical and the exact solution in the two-particle sector 
for the quantum model indicate that this limit is useful for the {\sl first} crossover, though.
% This is used in the next section for the analysis of the Lieb-Liniger equations for the attractive ground state.
% 
% 
% 
% 
% 
% 
% 
% 
% 
% 
% 
\subsubsection{Analysis of Lieb-Liniger equations for large $N$}
\label{sec:LL-BA-TBA}
The complicated form of the wave function within the coordinate Bethe Ansatz 
makes a straight forward extension of the ground-state wave function overlap
calculation similar to equation (\ref{eq:LL-overlap}) for the two-particle case impossible.
There exist determinant formulations via the Algebraic Bethe Ansatz~\cite{Korepins-book}.
These have been studied for the repulsive case only, though.

In the general Lieb-Liniger equations
the first excited energy is relatively complicated to treat, as the root pattern is not as simple 
as in the $N=2$ case. It can be shown that the roots never merge into the 
true $N$-string for $N>2$ for total momentum one, as this would violate hermiticity~\footnote{
Here $N$-string denotes $N$ Bethe ansatz roots with identical real part and symmetric imaginary part. In the literature
{\sl string} sometimes refers to the special case of uniform spacing.}.
The ground-state root configuration is more accessible, as it is generally believed to be 
an ideal $N$-string: the roots are purely imaginary and distributed symmetrically around the origin.
The limit of strong interaction or very large box size $L$ has been previously studied:
the  roots are then asymptotically linear in the interaction~\cite{Takahashi-book}
and evenly spaced. In this limit the (not normalised) wave function is of the McGuire
form~\cite{McGuire-attractive-scattering,Mattis-red-book}
\begin{align}
\Psi(x)= \exp\left({-\frac{|\gamma|}{2}  \sum_{i<j} |x_i -x_j|}\right)
\end{align}
which is also relevant to (infinite length) optical wave\-guides~\cite{Lai-waveguide}.
\begin {figure}[tp]
\begin{minipage}{\linewidth}
\begin{center}
\includegraphics[angle=-90,width=0.75\linewidth]{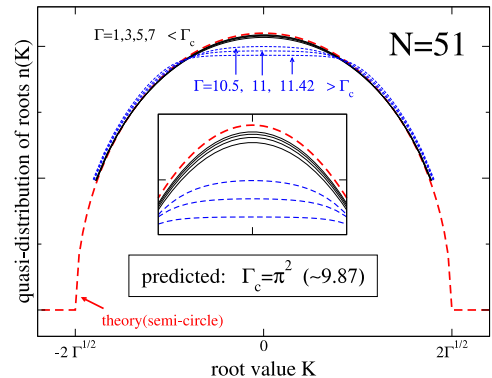}
\end{center}
\end{minipage}
 \caption{
	Quasi-distribution of the Bethe ansatz roots $n(K)$ for numerical solution of~(\ref{eq:generic-LL-BAE-K})
	with interaction $\Gamma$
	below the pre-transition value $\Gamma_c=\pi^2$.
	The roots follow the semi-circle law~(\ref{eq:SemiCircleLaw}), found analytically in the weak coupling limit 
	\mbox{$\Gamma\to0$}. The numerical results for $N=51$ are in agreement with the notion that the semi-circle law holds
	asymptotically for $\Gamma<\Gamma_c$ for sufficiently large $N$, while for $\Gamma>\Gamma_c$ the distribution 
	is uniform. The dashed blue lines show the start of the finite size crossover from the semi-circle towards
	the box shape~(\ref{eq:BoxLaw}). The inset is an enlargement of the center region where the inner-lying roots
	approach the uniform density first.
 		}
 \label{fig:LLdensity-K}
\end{figure}

To analyse the attractive ground state of an abstract system of equations of Lieb-Liniger type 
we introduce the real variables $K_j$ via $k_j \to \mathrm{i} K_j$. 
It is ad-hoc assumed that for {\sl finite} interaction and {\sl finite} $N$ there exists a unique
real solution to
\begin{align}
\mathrm{e}^{K_i}
= 
\prod_{j\neq i}^N
% \nonumber
\frac{
K_i - K_j + \frac{\Gamma}{N}
}{
K_i - K_j - \frac{\Gamma}{N}
},
\qquad  i=1...N\,.
\label{eq:generic-LL-BAE-K}
\end{align}
Here $\Gamma=c N $ is the rescaled interaction.
Note the similarity of these ground state equations to systems with hard wall boundary conditions~\cite{our-HW-paper,Gaudin-book}
due to the externally imposed symmetry of the roots.
The formulation of the problem in terms of the variables $\{K_i\}$ and $\Gamma$ allows the definition of a sensible distribution or quasi-density
of Bethe Ansatz roots for very large but finite particle numbers $N$.
Here it is useful to define a quasi-density of the roots $K_i$, which for example can be done via~\cite{our-mixture}
\begin{align}
n(x)=\left\{
\begin{tabular}{ll}
$\frac{1}{N-1}\frac{1}{K_{i+1}-K_i}$
&
\qquad$x\in(K_i,K_{i+1}]$
\\
$0$,		
&
\qquad$|x| > K_\textrm{max}$
\end{tabular}
\right.
% \nonumber
\label{eq:quasi-density}
\end{align}
In the weak coupling limit the root distribution of the real solution to~(\ref{eq:generic-LL-BAE-K})
follows a semi-circle law derived from the relation to the Hermite polynomials~\cite{Gaudin-book,weak-bosons}
\begin{align}
n(K) =\frac{1}{2\pi \Gamma} \sqrt{K^2_\textrm{max} \textendash K^2   } 
,\quad
|K|\leq K_\textrm{max}=2\sqrt{\Gamma}
\label{eq:SemiCircleLaw}
\end{align}
In the strong interaction limit the application of the string hypothesis leads to a uniform, box shaped density.
When constructing a string solution to~(\ref{eq:generic-LL-BAE-K})
for fixed $N$ and increasing $\Gamma\to\infty$ the difference
between closest roots is asymptotically
% $K_{i+1}-K_i = \frac{\Gamma}{N}$ 
$K_{i+1}-K_{i}=\frac{\Gamma}{N}$. Summing up over the symmetric root distribution
it follows that
$K_\textrm{max} = \frac{\Gamma}{2} \frac{N-1}{N}\to
\frac{\Gamma}{2}$,
%  $K_\textrm{max}=\frac{\Gamma}{2}$
this agrees with numerical exploration
for small particle numbers $N<50$.
\begin{align}
n(K) =\frac{1}{2 K_\textrm{max} }, \qquad
|K|\leq K_\textrm{max}=\frac12 {\Gamma}
\label{eq:BoxLaw}
\end{align}
The above expansions hold in the limits of $\Gamma\to0$ (weak) and $\Gamma\to\infty$ (strong), respectively, while $N$ (large) is
held constant.
Note that $K_\textrm{max}$ is in both cases
% , weak ($\Gamma\to0$) and strong ($\Gamma\to\infty$) interaction limit
{\sl independent} of the particle number $N$. This would allow, at least in principle, to explore the interesting limit $N\to\infty$
for a fixed and finite interaction strength $0<\Gamma<\infty$.
It is technically hard to relate the Bethe Ansatz roots in Lieb-Liniger type models to physical properties within the exact approach. 
Nevertheless, it is expected that the quasi-distribution~(\ref{eq:quasi-density})
of the Bethe Ansatz roots in~(\ref{eq:generic-LL-BAE-K}) for large $N$ (resp. $N\to\infty$)
will show qualitatively different behaviour in the two regions $\Gamma<\Gamma_c$ and $\Gamma>\Gamma_c$, see~\fref{fig:LLdensity-K}
for numerical results for $N=51$.

For weak interaction $\Gamma\ll\Gamma_c$ the numerical solution $\{K_i \}$ is distributed approximately as a
semi-circle~(\ref{eq:SemiCircleLaw}),
while for $\Gamma\gg\Gamma_c$ the quasi-density approaches a uniform box shape~(\ref{eq:BoxLaw}).
Numerical results
for small system sizes suggest an agreement with the expected value $\Gamma_c=\pi^2$ separating the two regions,
where $\Gamma_c$ is the location of the single minima in the ground state wave function overlap in the continuum model as discussed
in the earlier sections of this paper.
% 
% 
% 
% 
% 

% 
% 
% 
% 
% 
% 
% 
% ---------------------------------------------------------------------

Numerical solutions to finite $N$ Lieb-Liniger equations are usually found by starting with an initial
guess of the roots in a known region, e.g. the weak coupling limit. Then the interaction is
increased in small steps  $\Gamma \to \Gamma + \Delta$ where $N$ remains necessarily constant.
This so called {\sl root tracking} works well if the root set $\{k_i\}|_{\Gamma+\Delta}$ is similar to the previous step
$\{k_i\}|_{\Gamma}$ - for a close initial guess most non-linear solver have good convergence.
From the above it can be seen that this method is unsuitable for the study of large $N$ behavior - the root
pattern is expected to change strongly when crossing over the pre-transition at $\Gamma_c=\pi^2$, which is
in accordance with findings of Sakman et al.~\cite{attractive-integrable-LL-model}. In a diagram
$N$ vs. $\Gamma$ the above method corresponds to moving along horizontal lines, where in the left part
the root distribution is asymptotically of semi-circle shape, while on the right side it has the uniform box shape.
Using that the quasi-density (\ref{eq:quasi-density}) for finite particles is in one-to-one correspondence with
the root set $\{K_i \}$ the system (\ref{eq:generic-LL-BAE-K}) can be solved on {vertical lines}, i.e. for
fixed interaction $\Gamma$ and increasing $N$. In that way the solution is stable, i.e. it does not change significantly
for increasing $N$ as the pre-transition point is not crossed.

The preliminary numerical results obtained by root tracking agree with the behavior described above.
Nevertheless, a rigorous analysis of~(\ref{eq:generic-LL-BAE-K}) is 
necessary to determine  if a quantum phase transition occurs.
In particular the critical value $\Gamma_c=\pi^2$ has not been 
found from the Bethe Ansatz equations.
% , but was approximated by numerical solutions for small $N$.
This result will be relevant for the description of the first pre-transition of the initial finite size Bose-Hubbard model,
when transforming the considered abstract system back to the physical problem.
\section{CONCLUSIONS}
\label{sec:Discussion}
{
In this paper we have argued that there are signs of transitional behavior
in the ground state of the attractive one-dimensional Bose-Hubbard model. 
A discussion using conventional Quantum Phase transitions - defined in the thermodynamic
limit of many particles $N$ on many sites $L$ - is unsuitable as 
the standard limit for attractive bosons is subject to instant collapse.
Instead we have used the notion of pre-transitions, characterised by a sudden change in 
the ground-state properties when crossing a threshold interaction strength
in a system of large but finite size $N$, $L$.

Such pre-transitions are visible in indicator properties as
for instance the energy gap between ground state and 1st excitation
indicating onset of degeneracy, and local minima in the incremental overlap
$\langle \tau+\Delta|\tau\rangle$, where $|\tau\rangle$ is the ground state
for attractive interaction strength $\tau$.

We have used mean-field like approximations and integrable limits of the
model to examine regions inaccessible to exact diagonalisation, and compared with
exact numerics where applicable.
The transitional region depends on both lattice size $L$ and
number of bosons $N$ in a non-trivial way.
For specific parametrisations of the coupling
strength between the kinetic and the interaction contributions in the Hamiltonian one of the crossover points
is quasi stationary while the other wanders.
In particular we have shown that in the limit of very small and very large lattice size $L$,
the complex transitional regime reduces to only two
regimes with one single crossover point, in agreement with earlier studies on these models.
}
{
% \subsubsection{Experimental relevance}
% 
% 
The ground state is predicted to change strongly in a small region around critical attractive interactions.
In experiments with controlled change of attractive interaction this should have clearly visible effects 
in properties like correlation functions and momentum distribution.
If ultracold quantum gases with large but finite particle number $N$ and lattice size $L$,
enter the strong attractive interaction region the validity of 
the physical description by the simple Bose-Hubbard model needs to be carefully investigated, though.
}
In addition it will be interesting to see how this transitional behavior manifests in theories of more complex attractive boson systems, as 
we believe this is a generic feature of attractive bosonic systems rather than a speciality of this particular model.
The generalisations already studied in the repulsive regime like long-range hopping, long-range interactions
and extensions of lattice geometry to ladders and square lattices are an obvious starting point for further exploration.
For these systems there are currently few methods available using integrable techniques.
\\
% \vspace{.2cm}
% 
}
\section*{Acknowledgements}
The work was funded by the Australian Research Council under Discovery Project
DP0663773.
We thank
X-W Guan, M Bortz,  M T Batchelor and A Sykes for helpful discussions.
% 
% 
% \newpage

\end{document}